\documentclass[aps,prl,twocolumn,showpacs,amsfonts,superscriptaddress]{revtex4}

\usepackage{epsfig}
\usepackage{amsmath}
\usepackage{amssymb}
\usepackage{bm}

\newcommand{\beq}{\begin{equation}}
\newcommand{\eeq}{\end{equation}}
\newcommand{\beqarray}{\begin{eqnarray}}
\newcommand{\eeqarray}{\end{eqnarray}}

\newcommand{\Ham}[1][]{\ensuremath{{\cal{H}}_{\text{\tiny{#1}}}}} 
\newcommand{\eq}[1]{Eq.~(\ref{#1})} 
\newcommand{\fig}[1]{Fig.~(\ref{#1})} 
\newcommand{\Ref}[1]{Ref.~\onlinecite{#1}} 

\begin{document}

\allowdisplaybreaks

\title{$0$-$\pi$ transition in magnetic triplet superconductor
  Josephson junctions} 
\author{P. M. R. Brydon}
\email{brydon@theory.phy.tu-dresden.de}
\affiliation{Max-Planck-Institut f\"{u}r
  Festk\"{o}rperforschung, Heisenbergstr. 1, 70569 Stuttgart, Germany}
\affiliation{Institut f\"{u}r Theoretische Physik, Technische Universit\"{a}t  
  Dresden, 01062 Dresden, Germany }
\author{Dirk Manske}
\affiliation{Max-Planck-Institut f\"{u}r
  Festk\"{o}rperforschung, Heisenbergstr. 1, 70569 Stuttgart, Germany}

\date{\today}

\begin{abstract}
We examine a Josephson junction involving two arbitrary equal-spin-pairing
unitary triplet superconductors and a ferromagnetic tunneling barrier.
Using perturbation theory, we show how the interaction of the
barrier moment with the spin of the tunneling
triplet Cooper pairs can reverse the sign of the Josephson charge
current. This also results in a Josephson spin current, which
contains a phase-independent
contribution due to reflection
processes at the barrier. We verify our analytic predictions using a
non-perturbative Bogoliubov-de Gennes method.
\end{abstract}


\maketitle

{\it Introduction}. The interplay of superconductivity and magnetism is an
enduring 
enigma of condensed matter physics. %
Over the last 25 years, 
many fascinating insights into this problem have been
made in the study of singlet superconductor (SC) Josephson
junctions with ferromagnetic (FM) tunneling barriers~\cite{BegeretRMP2005}. For
example, as the barrier width is 
increased, the usual Josephson current $I_{J}$ vs phase relationship 
$I_{J}=|I_{0}|\sin(\phi)$ becomes $I_{J}=|I_{\pi}|\sin(\phi+\pi)$. This 
so-called $0$-$\pi$ transition is evidence of oscillations of the
singlet SC correlations in the tunneling
region~\cite{Buzdin1982,Ryazanov2001}. 
A remarkable feature of such junctions is the presence of triplet 
SC (TSC) correlations induced by a proximity
effect~\cite{BegeretRMP2005,Eschrig2003,Yokoyama}, with the
realized triplet pairing  
states dictated by the details of the FM barrier and the bulk
SCs. Due to the likely intimate connection between triplet
superconductivity and magnetism~\cite{Anderson1984}, it is interesting to
consider the case 
where the TSC pairing state can be chosen independently of the FM barrier.
Despite the growing
interest~\cite{Yokoyama,spin,KasteningTFT2006,BrydonTFT2008,Brydonspin2008}
in TSC Josephson junctions prompted by the discovery of
Sr$_2$RuO$_4$~\cite{Sr2RuO4}, the study of such TSC--FM--TSC (TFT) junctions is
still in its infancy. Recently, a novel 
$0$-$\pi$ transition in a specific TFT junction was 
predicted~\cite{KasteningTFT2006,BrydonTFT2008}, where the dependence of
$I_{J}$ upon the \emph{orientation} of the FM moment indicates
that it couples to the spin of the tunneling Cooper pairs.  

In this letter, we use perturbation
theory~\cite{Mahan} to obtain the 
Josephson charge 
current through a TFT junction for arbitrary choice of unitary
equal-spin-pairing TSCs. We predict that the $0$-$\pi$ transition
found in Ref.s~\onlinecite{KasteningTFT2006}
and~\onlinecite{BrydonTFT2008} is always present for sufficiently large
magnetization, and is due to the spin-flipping of tunneling triplet Cooper
pairs. This also produces a
Josephson spin current~\cite{spin,Brydonspin2008}, which has opposite sign on
either side of the junction and a novel phase-independent 
contribution
due to reflection processes. A non-perturbative Bogoliubov-de Gennes
theory is used to demonstrate the universal character of our predictions, and
that 
resonant tunneling through an Andreev 
bound state (ABS) does not qualitatively change the understanding of
the $0$-$\pi$ 
transition~\cite{Fogelstroem2000,Kashiwaya2000}. 

{\it Perturbation theory}. The Hamiltonian describing the TFT Josephson
junction is written 
$\Ham = \Ham[L] + \Ham[R] + \Ham[tun] + \Ham[ref]$. The terms $\Ham[L]$ and
$\Ham[R]$ respectively describe the bulk TSCs on the left and right side of
the barrier: 
\beq
{\cal{H}}_{\nu} =
\frac{1}{2}\sum_{\bf{k}}\psi^{\dagger}_{\nu,\bf{k}}\left(\begin{array}{cc}
\epsilon_{\nu,\bf{k}}\hat{1} &
i{\bf{d}}_{\nu,{\bf{k}}}\cdot{\pmb{\sigma}}\hat{\sigma}_{y} \\
(i{\bf{d}}_{\nu,{\bf{k}}}\cdot{\pmb{\sigma}}\hat{\sigma}_{y})^{\dagger} &
-\epsilon_{\nu,{\bf{k}}}\hat{1}
\end{array}
\right)\psi^{}_{\nu,\bf{k}}
\eeq
where $\psi^{}_{\nu,\bf{k}}=(c^{}_{\nu,{\bf{k}},\uparrow},c^{}_{\nu,{\bf{k}},\downarrow},c^{\dagger}_{\nu,{\bf{k}},\uparrow},c^{\dagger}_{\nu,{\bf{k}},\downarrow})^{T}$ and $c^{\dagger}_{\nu,{\bf{k}},\sigma}$ ($c^{}_{\nu,{\bf{k}},\sigma}$)
are fermion creation (annihilation) operators,
$\epsilon_{\nu,{\bf{k}}}$ is the bare 
dispersion in the $\nu$-hand TSC, and
${\bf{d}}_{\nu,{\bf{k}}}=\Delta_{\nu,{\bf{k}}}\hat{\bf{x}}$ are the triplet
order parameters of the two TSCs. Both TSCs are in
an equal spin-pairing state with respect to the $z$-axis and are unitary
(i.e. the triplet condensate has no net spin)~\cite{other}. 
The gap in each spin sector is 
$\Delta_{\nu,{\bf{k}},\sigma}=-\sigma|\Delta_{\nu,{\bf{k}}}|e^{i(\phi_{\nu}+\theta_{\nu,{\bf{k}}})}$
where $\phi_{\nu}$ is the global phase of the $\nu$-hand TSC
and $\theta_{\nu,{\bf{k}}}$ is an internal phase specifying the 
pairing state,  
obeying $\theta_{\nu,-{\bf{k}}}=\theta_{\nu,{\bf{k}}}+\pi$.
As our results depend \emph{only} on the spin state of the Cooper pairs, any 
variation of the orbital part of the gaps near the barrier will not
qualitatively alter our conclusions.

The two TSCs on each side of the barrier are linked by the
tunneling Hamiltonian  
\beq
\Ham[tun] =
\sum_{\nu=L,R}\sum_{{\bf{k}},{\bf{k}}'}\sum_{\sigma,\sigma'}T^{\sigma,\sigma'}_{\nu,{\bf{k}},{\bf{k}}'}c^{\dagger}_{-\nu,{\bf{k}},\sigma}c^{}_{\nu,{\bf{k}}',\sigma'}
\eeq
where
the subscript $-\nu=R(L)$ when $\nu=L(R)$. 
For a magnetically-active barrier, we must also include reflection
prcoesses~\cite{Cuevas2001}: 
\beq
\Ham[ref] =
\sum_{\nu=L,R}\sum_{{\bf{k}},{\bf{k}}'}\sum_{\sigma,\sigma'}R^{\sigma,\sigma'}_{\nu,{\bf{k}},{\bf{k}}'}c^{\dagger}_{\nu,{\bf{k}},\sigma}c^{}_{\nu,{\bf{k}}',\sigma'}
 \label{eq:Hamref}
\eeq
Althogh reflection processes do not contribute to the charge current,
{\it spin-flip} reflection may
contribute to a Josephson spin current, as the spin-flip of a
reflected Cooper pair changes the total spin in the TSC by
$\pm2\hbar$. 

\begin{figure}
  \includegraphics[width=1.01\columnwidth]{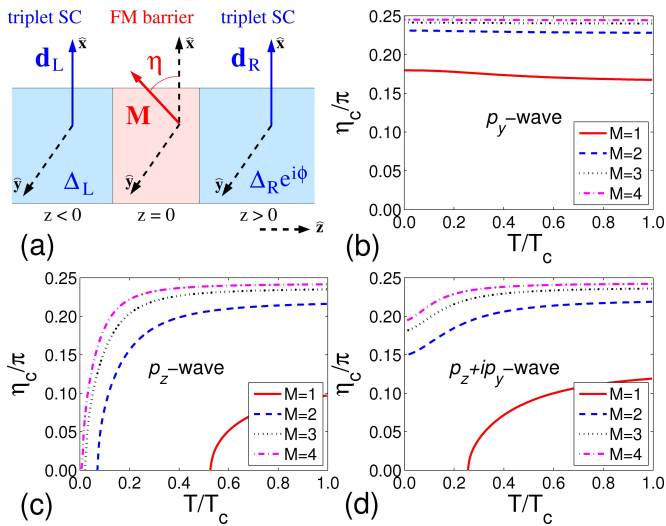}
  \caption{\label{etac} (color online) (a) Schematic diagram of the TFT
    junction. (b-d) The critical line $\eta_{c}$ between the $0$ 
    and $\pi$ 
    states for various values of $M$ in the (b) $p_{y}$, (c) $p_{z}$, and (d)
    $p_{z}+ip_{y}$ junctions. The $\pi$ state is 
    realized for $0\leq|\eta|,|\pi-\eta|<\eta_{c}$, while the $0$ state occurs
    for $\eta_{c}<|\eta|<\pi-\eta_{c}$. The $M\gg{1}$ behaviour of
    $\eta_{c}$ at high $T$ reflects the dominance of spin-flip processes; at
    low $T$, details of the bulk TSCs are important.}   
\end{figure}

In general, the matrix elements for spin-preserving
tunneling $T^{\sigma,\sigma}_{\nu,{\bf{k}},{\bf{k}}'}$, spin-flip
tunneling $T^{\sigma,-\sigma}_{\nu,{\bf{k}},{\bf{k}}'}$, and spin-flip
reflection $R^{\sigma,-\sigma}_{\nu,{\bf{k}},{\bf{k}}'}$ are
different. It is possible to derive expressions for the matrix
elements from a more fundamental Hamiltonian~\cite{Cuevas2001}, but here we 
will motivate a phenomenological form. By Fermi's golden
rule we have  
${\cal{T}}^{\sigma,\sigma'}\sim|T^{\sigma,\sigma'}_{\nu,{\bf{k}},{\bf{k}}'}|^2$
and
${\cal{R}}^{\sigma,-\sigma}\sim|R^{\sigma,-\sigma}_{\nu,{\bf{k}},{\bf{k}}'}|^2$
in the tunneling limit,
where ${\cal{T}}^{\sigma,\sigma'}\ll{1}$ and
${\cal{R}}^{\sigma,-\sigma}\ll{1}$ are the 
transmissivity and spin-flip reflectivity of the barrier respectively. We also
require 
that a tunneling or reflected quasiparticle acquires the same phase as in the
exact solution. Following~\Ref{KasteningTFT2006}, we consider the example of a
purely FM barrier of  
$\delta$-function width (appropriate for an atomically-thin barrier) at
$z=0$. We assume that the FM barrier moment ${\bf{M}}$
lies 
in the $x$-$y$ plane at an angle $\eta$ to the $x$-axis. We hence use the Ansatz
\begin{align}
T^{\sigma,\sigma}_{\nu,{\bf{k},{\bf{k}}}'} &=
(T_{sp}/M^2)\delta_{{\bf{k}}_{\parallel},{\bf{k}}'_{\parallel}}\theta(k_{z}k'_{z}) \label{eq:Tss} 
\\
T^{\sigma,-\sigma}_{\nu,{\bf{k},{\bf{k}}'}} &=
-\nu{i}e^{-i\sigma\eta}(T_{sf}/M)\delta_{{\bf{k}}_{\parallel},{\bf{k}}'_{\parallel}}\theta(k_{z}k'_{z}) \label{eq:Ts-s}\\  
R^{\sigma,-\sigma}_{\nu,{\bf{k}},{\bf{k}}'} &=
\nu{i}e^{-i\sigma\eta}(R_{sf}/M)\delta_{{\bf{k}}_\parallel,{\bf{k}}'_{\parallel}}\theta(-k_{z}k'_{z}) \label{eq:Rs-s} 
\end{align}
where $M=g\mu_{B}|{\bf{M}}|/\hbar^2\sqrt{v^{L}_{F,z}v^{R}_{F,z}}$ with
$v^{\nu}_{F,z}$ the  
Fermi velocity along the $(001)$-direction in the $\nu$-hand
TSC, 
$T_{sp}$, $T_{sf}$ and $R_{sf}$ are real constants 
and the $\theta(\pm{k_{z}k_{z}'})$ guarantees that the transmitted or reflected
quasiparticle moves away from the barrier~\cite{Bruder1995}. The
${\bf{k}}$-dependence of $T_{sp}$, $T_{sf}$ and $R_{sf}$ is irrelevant for our
argument and is neglected. 
As rotating the spin coordinates about the $x$ axis leaves the TSCs
unchanged, our results for the charge current hold for any moment making an
angle $\eta$ with the ${\bf d}_{\nu,{\bf{k}}}$ vectors. 
The spin current results also hold, but with 
corresponding rotation of the polarization. A schematic diagram of the
junction is shown in~\fig{etac}(a).

We define particle currents in the two spin sectors of each TSC by
$I_{\nu,\alpha}=-\nu\langle\partial_{t}N_{\nu,\alpha}(t)\rangle$ where
$N_{\nu,\alpha}(t)=\sum_{\bf{k}}c^{\dagger}_{\nu,{\bf{k}},\alpha}(t)c^{}_{\nu,{\bf{k}},\alpha}(t)$
and $\nu=L(R)$ as a subscript implies $\nu=-1(+1)$ elsewhere. We calculate
$I_{\nu,\alpha}$ by expanding the $S$
matrix to lowest order in $\Ham[tun]+\Ham[ref]$, hence treating the
tunneling and reflection processes as a perturbation of
${\cal{H}}_{0}=\Ham[L]+\Ham[R]$~\cite{Mahan}, which is justified for small  
$T^{\sigma,\sigma'}_{\nu,{\bf{k}},{\bf{k}}'}$ and
$R^{\sigma,\sigma'}_{\nu,{\bf{k}},{\bf{k}}'}$. The Kubo
formula then gives 
$
I_{\nu,\alpha}=-i\nu\int^{t}_{-\infty}dt'\langle\left[\partial_{t}N_{\nu,\alpha}(t),\Ham[tun](t')
  + \Ham[ref](t')\right]\rangle$. Working
within the interaction picture, we write
$\partial_{t}N_{\nu,\alpha}(t)=i\left\{B^{-\alpha,\alpha}_{\nu}(t) -
  B^{\alpha,-\alpha}_{\nu}(t)\right\} +
i\sum_{\sigma}\left\{A^{\sigma,\alpha}_{\nu}(t) -
  A^{\alpha,\sigma}_{-\nu}(t)\right\}$ where we define the operators 
$A^{\varsigma,\varsigma'}_{\nu}(t)=\sum_{\bf{k},\bf{k}'}T^{\varsigma,\varsigma'}_{\nu,\bf{k},\bf{k}'}c^{\dagger}_{-\nu,{\bf{k}},\varsigma}(t)c^{}_{\nu,{\bf{k}}',\varsigma'}(t)$
and $B^{\varsigma,\varsigma'}_{\nu}(t)=\sum_{\bf{k},\bf{k}'}R^{\varsigma,\varsigma'}_{\nu,{\bf{k}},{\bf{k'}}}c^{\dagger}_{\nu,{\bf{k}},\varsigma}(t)c^{}_{\nu,{\bf{k}}',\varsigma'}(t)$.  
We consider only the zero bias case, where the
current through the spin-$\alpha$ sector of the $\nu$-hand
TSC is due only to the Josephson effect and may be conveniently
expressed as
$I^{J}_{\nu,\alpha} =
2\nu\mbox{Im}\left\{\Phi^{r}_{\nu,\alpha}(0) +
\Psi^{r}_{\nu,\alpha}(0)\right\}$. 
The retarded correlation functions $\Phi^{r}_{\nu,\alpha}(\omega)$ and
$\Psi^{r}_{\nu,\alpha}(\omega)$ give the tunneling and
reflection contributions respectively. After using Wick's theorem to expand the
two-particle correlators in the corresponding Matsubara functions
$\Phi_{\nu,\alpha}(i\omega_n) = \int^{\beta}_{0} d\tau
e^{i\omega_n{\tau}}\sum_{\sigma,s,s'} \langle{T_\tau
    A^{\sigma,\alpha}_{\nu}(\tau)A^{s,s'}_{\nu}(0)}\rangle $ and $\Psi_{\nu,\alpha}(i\omega_n) = \int^{\beta}_{0} d\tau
e^{i\omega_n{\tau}}\left\langle{T_\tau
    B^{-\alpha,\alpha}_{\nu}(\tau)B^{-\alpha,\alpha}_{\nu}(0)}\right\rangle$,
we make the analytic continuation $i\omega_{n}\rightarrow \omega +
i0^{+}$ to obtain the retarded functions. 
Substituting~Eq.s~(\ref{eq:Tss}-\ref{eq:Rs-s}) into the expressions for the
retarded correlation functions at $\omega=0$, we obtain the 
particle current
\begin{align}
I^{J}_{\nu,\alpha} =& -\sum_{{\bf{k}},{\bf{k}}'}\frac{T_{sp}^2}{M^4}\frac{|\Delta_{-\nu,{\bf{k}}}\Delta_{\nu,{\bf{k}}'}|}{E_{-\nu,{\bf{k}}}E_{\nu,{\bf{k}}'}}F_{-\nu,\nu,{\bf{k}},{\bf{k}}'}\delta_{{\bf{k}}_{\parallel},{\bf{k}}'_{\parallel}}\theta(k_{z}k'_{z}) \notag \\
&\qquad\times\sin(\phi+\nu[\theta_{\nu,{\bf{k}}'}-\theta_{-\nu,{\bf{k}}}])
\notag \\
& + \sum_{{\bf{k}},{\bf{k}}'}\frac{T_{sf}^2}{M^2}\frac{|\Delta_{-\nu,{\bf{k}}}\Delta_{\nu,{\bf{k}}'}|}{E_{-\nu,{\bf{k}}}E_{\nu,{\bf{k}}'}}
F_{-\nu,\nu,{\bf{k}},{\bf{k}}'}\delta_{{\bf{k}}_{\parallel},{\bf{k}}'_{\parallel}}\theta(k_{z}k'_{z})   \notag \\
& \qquad\times \sin(\phi +
2\nu\alpha\eta+\nu[\theta_{\nu,{\bf{k}}'}-\theta_{-\nu,{\bf{k}}}]) \notag \\
& + \nu\sum_{{\bf{k}},{\bf{k}}'}\frac{R_{sf}^2}{M^2} \frac{|\Delta_{\nu,{\bf{k}}}\Delta_{\nu,{\bf{k}}'}|}{E_{\nu,{\bf{k}}}E_{\nu,{\bf{k}}'}}
F_{\nu,\nu,{\bf{k}},{\bf{k}}'}\delta_{{\bf{k}}_{\parallel},{\bf{k}}'_{\parallel}}\theta(-k_{z}k'_{z}) \notag\\
& \qquad\times \sin(2\alpha\eta-[\theta_{\nu,{\bf{k}}} -
\theta_{\nu,{\bf{k}}'}]) \label{eq:IJnafull}
\end{align}
where $E_{\nu,{\bf{k}}}=\sqrt{(\epsilon_{\nu,{\bf{k}}}-\mu)^2+|\Delta_{\nu,{\bf{k}}}|^2}$ 
is the excitation spectrum in the $\nu$-hand TSC,
$\phi=\phi_{R}-\phi_{L}$ and $F_{\nu,\nu',{\bf{k}},{\bf{k}}'} =
\sum_{\pm}[n_{F}(\pm{E_{\nu,{\bf{k}}}}) - 
  n_{F}(E_{\nu',{\bf{k}}'})]/[E_{\nu,{\bf{k}}} \mp
  E_{\nu',{\bf{k}}'}]$ with 
$n_{F}(E)$ the Fermi distribution function.

\eq{eq:IJnafull} is our first
important result, as it contains
all contributions to the current. The first term describes
spin-preserving tunneling, where the Cooper pairs preserve their spin during
the tunneling event, giving the usual Josephson result. The second term
describes spin-flip tunneling, where the 
spin of the Cooper pair is reversed by the coupling to the FM moment. Relative
to spin-preserving tunneling, these Cooper pairs acquire a
phase shift of 
$2\alpha\nu\eta$ due to the spin-flip itself, and a further $\pi$-shift
arising from the intrinsic phase difference between the spin-$\uparrow$ and
spin-$\downarrow$ condensates in the TSC. Lastly, we have the current due to 
Cooper pairs undergoing a spin-flip when they are reflected at the tunneling
barrier. As such, this term is independent of the TSC on the other side of
the barrier, depending only upon the phase due to the spin-flip itself and
the gap experienced by the reflected Cooper pairs. 

{\it Charge current}. From~\eq{eq:IJnafull} we obtain the Josephson charge
current $I_{J}=-e(I^J_{\nu,\uparrow}+I^{J}_{\nu,\downarrow})$: 
\begin{align}
I_{J} =& 2e\left(\frac{T_{sp}^2}{M^4} - \cos(2\eta)\frac{T_{sf}^2}{M^2}\right)\sum_{{\bf{k}},{\bf{k}}'}\frac{|\Delta_{R,{\bf{k}}}\Delta_{L,{\bf{k}}'}|}{E_{R,{\bf{k}}}E_{L,{\bf{k}}'}}F_{R,L,{\bf{k}},{\bf{k}}'} \notag \\
& \times \delta_{{\bf{k}}_{\parallel},{\bf{k}}'_{\parallel}}\theta(k_{z}k'_{z}) \sin(\phi+\theta_{R,{\bf{k}}}-\theta_{L,{\bf{k}}'})
 \label{eq:totalcur}
\end{align}
The first term in brackets corresponds to the spin-preserving contribution,
while the second term is due to the spin-flip
tunneling. The $\eta$-dependence of the latter is due to the extra phase
shifts for  
spin-flip tunneling: ignoring orbital effects, a spin-$\alpha$ 
Cooper pair incident from the LHS undergoing a spin-flip during tunneling
experiences an effective  
phase difference $\phi_{\alpha}=\phi+\pi-2\alpha\eta$ between
the two TSCs. The spin-flip current vs phase
relationships in each spin channel are hence shifted by $\pm4\eta$ with
respect to one another. The interference between the two spin
channels results 
in the modulation of the total spin-flip current by 
$\cos(2\eta)$; this is analogous to the effect of the spin-dependent phase
shifts for tunneling between TSCs with misaligned
${\bf{d}}$-vectors~\cite{spin,KasteningTFT2006,BrydonTFT2008,Brydonspin2008}.

Spin-flip tunneling dominates $I_{J}$ for large $M$: in this case,
relevant for a half-metallic barrier, we find $I_{J}\propto\cos(2\eta)$, and
so the current changes sign at $\eta=\pi/4$ as the moment is rotated about the
$z$-axis, i.e. there is a $0$-$\pi$ transition controlled by the orientation
of the moment. 
As this originates solely from the spin structure of the triplet Cooper pairs
it is an {\it universal} feature of unitary equal-spin-pairing TFT
junctions, our second important result.
To test this prediction, we consider TFT junctions where both TSCs are
made of the same material, for the three choices of 
$p_{y}$, $p_{z}$, and 
$p_{z}+i{p_{y}}$ orbital pairing symmetry (the latter of most relevance to
Sr$_{2}$RuO$_{4}$~\cite{Sr2RuO4}). In a model with spatially constant TSC
gaps, we can include tunneling and reflection
processes to all orders~\cite{Kashiwaya2000,Cuevas2001,Zhao2004} by solving the
Bogoliubov-de Gennes equations to obtain the ABS energies 
  $E_{\pm,{\bf{k}}}(\phi,\eta,M,T)$ at 
temperature 
$T$. The free 
energy of the junction is then given by $F =
-\beta\sum_{\bf{k}}\sum_{p=\pm}(|k_{z}|/k_{F})\log(2\cosh({\beta}E_{p,{\bf{k}}}/2))$;
we assume a 2D circular Fermi surface in the $y$-$z$ plane. 
The ground state of the junction is found by numerically minimizing $F$
with respect to 
$\phi$. For each junction, the minimum of $F$ lies at
$\phi=0$ or $\pi$; we find a $0$-$\pi$ transition when the 
global minimum 
shifts from one of these values to the other. In~\fig{etac}(b-d) we plot the
critical 
angle $\eta_{c}$ at which this occurs in each junction for fixed $M$; 
according to~\eq{eq:totalcur}, in the tunneling limit $\eta_{c}=\pi/4$. 

For $T$ sufficiently close to the transition
temperature $T_c$, $\eta_c$ always approaches the tunneling
limit results as $M$ is increased. For the $p_{y}$ junction
[\fig{etac}(b)], $\eta_{c}$ shows only weak $T$-dependence at fixed $M$,
consistent with~\eq{eq:totalcur}.
In the $p_z$ and $p_{z}+ip_{y}$ junctions [\fig{etac}(c) and (d)
  respectively], however, $\eta_{c}$ varies significantly with $T$. 
The key difference between the $p_{y}$ junction and the $p_{z}$ and
$p_{z}+ip_{y}$ junctions is the absence of a zero energy ABS in
the former. In the latter, there is a zero energy ABS at $\phi=\pi$ for
\emph{any} choice of tunneling 
barrier, 
which raises the free energy of the $\pi$ state, thereby suppressing the
$0$-$\pi$ transition. 
The strongest
deviations from~\eq{eq:totalcur} therefore occur for the
$p_z$ junction, as here a zero energy 
ABS forms for all ${\bf{k}}$; for the $p_{z}+ip_{y}$ junction, in
contrast, a zero energy ABS forms only when ${\bf{k}}\parallel\hat{\bf{z}}$.
Our perturbation theory results are recovered at higher $T$ in these
junctions due to the suppression of multiple-Cooper pair tunneling processes,
which are a key feature of tunneling through a zero energy
ABS~\cite{Kashiwaya2000,resonant}.

\begin{figure}
\includegraphics[width=\columnwidth]{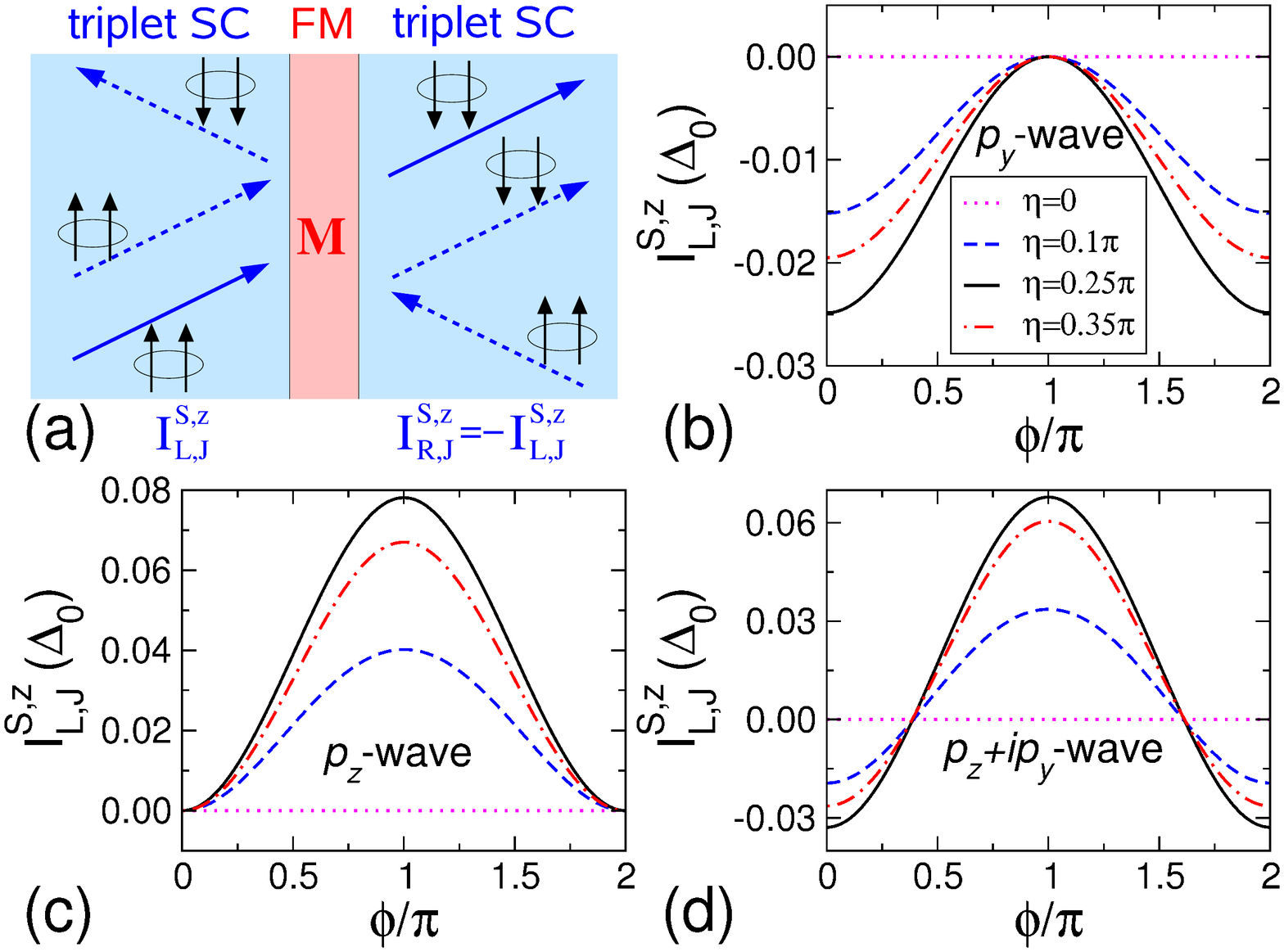}
\caption{\label{spincur} (color online) (a) Cartoon of spin-flip tunneling
  (solid line) and reflection (broken line) processes contributing to the spin
  current. (b-d) The $z$-component of spin current
  on the LHS for various values of $\eta$ at $M=2$ and $T=0.4T_{c}$ in the
  (b) $p_{y}$, (c) $p_{z}$, and (d) $p_{z}+ip_{y}$ junctions. The legend in
  (b) is for all plots. $\Delta_{0}$ is the maximum gap magnitude at
  $T=0$.}  
\end{figure}

{\it Spin current}. The spin of a triplet Cooper pair also allows a zero-bias
Josephson spin current to 
flow across the junction~\cite{spin,BrydonTFT2008,Brydonspin2008}.
The spin current is polarized along the $z$-axis~\cite{Brydonspin2008}, and
so from~\eq{eq:IJnafull} we obtain the spin 
current in the $\nu$-hand TSC
$I^{S,z}_{\nu,J}=\frac{\hbar}{2}(I^J_{\nu,\uparrow}-I^J_{\nu,\downarrow})$: 
\beqarray
I^{S,z}_{\nu,J} &=&
\hbar\nu\sin(2\eta)\sum_{{\bf{k}},{\bf{k}}'}\frac{T_{sf}^2}{M^2} \frac{|\Delta_{R,{\bf{k}}}\Delta_{L,{\bf{k}}'}|}{E_{R,{\bf{k}}}E_{L,{\bf{k}}'}}
   \notag \\
&&\times \delta_{{\bf{k}}_{\parallel},{\bf{k}}'_{\parallel}}\theta(k_{z}k'_{z})\cos(\phi+\theta_{R,{\bf{k}}}-\theta_{L,{\bf{k}}'})F_{R,L,{\bf{k}},{\bf{k}}'}  \notag \\
&& + \hbar\nu\sin(2\eta)
  \sum_{{\bf{k}},{\bf{k}}'}\frac{R^{2}_{sf}}{M^2}\frac{|\Delta_{\nu,{\bf{k}}}\Delta_{\nu,{\bf{k}}'}|}{E_{\nu,{\bf{k}}}E_{\nu,{\bf{k}}'}}
  \notag \\
&& \times \delta_{{\bf{k}}_{\parallel},{\bf{k}}'_{\parallel}}\theta(-k_{z}k'_{z})\cos(\theta_{\nu,{\bf{k}}} -
\theta_{\nu,{\bf{k}}'})
F_{\nu,\nu,{\bf{k}},{\bf{k}}'} \label{eq:spincur}
\eeqarray
The first term is from spin-flip
tunneling, while the second $\phi$-independent term is due to
spin-flip reflection~\cite{Grein2009}. The spin-dependent phase shifts of the
spin-flipping Cooper pairs are responsible for driving the spin current, again
in analogy to the spin current between TSCs with misaligned ${\bf{d}}$
vectors~\cite{spin,BrydonTFT2008,Brydonspin2008}. This
implies $I^{S,z}_{\nu,J}=0$ for 
$\eta=n\pi/2$, $n\in\mathbb{Z}$, as the relative phase between the
spin-flip currents in each spin channel is then $2n\pi$.
As shown in~\fig{spincur}(a), the spin current reverses sign
across the barrier: the spin 
current carried by the tunneling Cooper pairs reverses on the spin-flip, 
while the spin-flip reflected Cooper pairs in each TSC carry opposite spin
current as 
they move in opposite directions.

\eq{eq:spincur} may be simplified for the three junctions
introduced above. By energy conservation we have ${\bf{k}}'={\bf{k}}$ 
in the tunneling term and ${\bf{k}}'=\widetilde{\bf{k}}=(k_{x},k_{y},-k_{z})$
in the reflection term. Furthermore, we set
$T_{sf}=R_{sf}$ as here 
${\cal{T}}^{\sigma,-\sigma}={\cal{R}}^{\sigma,-\sigma}$; the amplitude of the
cosine term in each contribution 
to~\eq{eq:spincur} is 
then identical. We hence find $I^{S,z}_{\nu,J} \propto
\gamma + \cos(\phi)$ where $\gamma$ is an orbital-dependent
constant due to the phase shift
$\Delta\theta_{\nu,\bf{k}} = \theta_{\nu,{\bf{k}}}-\theta_{\nu,\widetilde{\bf{k}}}$ 
experienced by specularly-reflected Cooper pairs. 
For the $p_{y}$ junction we have
$\theta_{\nu,\widetilde{\bf{k}}}=\theta_{\nu,\bf{k}}$, and so
there is no extra phase shift upon reflection, giving $\gamma=1$. For
the $p_z$ junction, in contrast, all reflected Cooper pairs
experience a $\pi$ phase shift 
and therefore $\gamma=-1$. In the $p_{z}+ip_{y}$ junction,
$\Delta\theta_{\nu,{\bf{k}}}=\pi-2\arctan(k_{y}/k_{z})$ depends upon
${\bf{k}}$; integrating  
across the Fermi surface we find $-1<\gamma<0$. 
We again verify these predictions within the Bogoliubov-de Gennes theory
for spatially-constant TSC gaps. Solving
for the scattering
wavefunctions~\cite{BrydonTFT2008,Kashiwaya2000,Furusaki1991}, 
we obtain the Andreev reflection amplitudes 
$a^{eh(he)}_{\nu,\sigma,\sigma'}$ for a spin-$\sigma$ electron-like (hole-like)
quasiparticle incident from the $\nu$-hand-side Andreev-reflected as a
spin-$\sigma'$ hole-like (electron-like)
quasiparticle. Following~\Ref{Furusaki1991}, we write the spin current in
terms of the $a^{eh(he)}_{\nu,\sigma,\sigma'}$
\beqarray
I^{S,z}_{\nu,J} & = & -\frac{\nu}{8}\int_{|{\bf{k}}|=k_{F}}d{\bf{k}}\frac{|k_{z}|}{k_{F}}\frac{1}{\beta\hbar}\sum_{n}
\frac{|\Delta_{\nu,{\bf{k}}}|}{\sqrt{\omega^2_{n} +
    |\Delta_{\nu,{\bf{k}}}|^2}}\notag \\
&&\times\sum_{\sigma}\sigma\left\{a^{eh}_{\nu,\sigma,\sigma}({\bf{k}},i\omega_{n})-a^{he}_{\nu,\sigma,\sigma}({\bf{k}},i\omega_{n})\right\}
\eeqarray
We verify the relation $I^{S,z}_{L,J}=-I^{S,z}_{R,J}$ (not shown), and also
find excellent agreement with the tunneling Hamiltonian predictions for 
$\gamma$ in all three junctions, see~\fig{spincur}(b-d). The role played by
reflection processes in the spin transport is our third important result. 

We have not accounted for the transfer of spin to the barrier moment
when the spin current is non-zero. This can be physically justified if the
barrier is in contact with a spin reservoir, allowing the 
diffusion of the transferred spin. In the absence of
such a reservoir, we speculate that the barrier moment will precess about the
$x$-axis, as the spin current only has a spin polarization $\parallel{\bf{d}}\times{\bf{M}}$. This very interesting matter requires a 
non-equilibrium treatment, which is
beyond the scope of the present work. As $I^{S,z}_{\nu,J}=0$ for
$\eta=0$  and $\eta=\pi/2$, the different sign of $I_{J}$ at these
angles is however a robust equilibrium feature. 

{\it Conclusions.} We have analyzed the Josephson currents through a TFT
junction for any choice of unitary equal-spin-pairing TSCs. We predict
that the sign of the charge current is controlled 
by the relative importance of spin-flip to spin-preserving
tunneling. Spin-flip processes also produce a Josephson spin 
current, with a phase-independent term due to reflection. Our results reveal
the importance of the Cooper pair spin as a novel degree of freedom in
TSC Josephson junctions. 

We thank Y. Asano, J. Linder, D. K. Morr and
M. Sigrist for useful discussions, with special thanks to B. Rosenow and 
C. Timm.

\end{document}